\documentclass[conference, A4]{IEEEtran}
\IEEEoverridecommandlockouts

\usepackage{setspace}
\usepackage{float}
\usepackage{dblfloatfix}
\usepackage{cite}
\usepackage{amsmath,amssymb,amsfonts}
\usepackage{algorithm}
\usepackage{algpseudocode}
\usepackage{algorithmicx}
\usepackage{graphicx}
\usepackage{textcomp}
\usepackage{caption}
\usepackage{subcaption}
\usepackage{xcolor}

\usepackage{enumitem}
\usepackage[font=footnotesize]{caption}
\usepackage{tikz}

\usepackage{fancyhdr}
\newcommand*\circled[1]{\tikz[baseline=(char.base)]{
            \node[shape=circle,fill,inner sep=1pt] (char) {\textcolor{white}{#1}};}}

\makeatletter
\renewcommand\footnoterule{
  \kern-3\p@
  \hrule\@width 0.5\columnwidth
  \kern2.6\p@}
  \makeatother

\def\BibTeX{{\rm B\kern-.05em{\sc i\kern-.025em b}\kern-.08em
    T\kern-.1667em\lower.7ex\hbox{E}\kern-.125emX}}
    
\renewcommand{\IEEEbibitemsep}{0.3pt}
\makeatletter
\IEEEtriggercmd{\reset@font\normalfont\fontsize{7.2pt}{7.5pt}\selectfont}
\makeatother
\IEEEtriggeratref{1}

\begin{document}

\title{Reduce: A Framework for Reducing the Overheads of Fault-Aware Retraining}

\author{\IEEEauthorblockN{Muhammad Abdullah Hanif, Muhammad Shafique}
\IEEEauthorblockA{\textit{Division of Engineering, New York University
Abu Dhabi (NYUAD), Abu Dhabi, United Arab Emirates}\\
mh6117@nyu.edu, muhammad.shafique@nyu.edu}\vspace{-10mm}}

\maketitle
\thispagestyle{fancy}
\begin{abstract}
Fault-aware retraining has emerged as a prominent technique for mitigating permanent faults in Deep Neural Network (DNN) hardware accelerators. However, retraining leads to huge overheads, specifically when used for fine-tuning large DNNs designed for solving complex problems. Moreover, as each fabricated chip can have a distinct fault pattern, fault-aware retraining is required to be performed for each chip individually considering its unique fault map, which further aggravates the problem. To reduce the overall retraining cost, in this work, we introduce the concept of resilience-driven retraining amount selection. 
To realize this concept, we propose a novel framework, \textit{Reduce}, that, at first, computes the resilience of the given DNN to faults at different fault rates and with different amounts of retraining. Then, based on the resilience, it computes the amount of retraining required for each chip considering its unique fault map. 
We demonstrate the effectiveness of our methodology for a systolic array-based DNN accelerator experiencing permanent faults in the computational array. 
\end{abstract}

\section{Introduction}
\label{Sec1:Introduction}

Deep Neural Networks (DNNs) have emerged as a promising set of models for solving complex problems~\cite{lecun2015deep}. 
They are now state of the art for many AI applications, e.g., image classification, object detection and language translation~\cite{lecun2015deep}\cite{sze2017efficient}. 
However, these DNNs have high computational complexity~\cite{sze2017efficient}. 
To meet stringent performance and efficiency constraints of real-world applications, specialized DNN hardware accelerators, such as Eyeriss~\cite{chen2019eyeriss} and TPU~\cite{jouppi2017datacenter}, are used. These accelerators are usually built using nano-scale CMOS technology and face various reliability issues. 

One of the foremost reliability concerns with nano-scale CMOS devices is permanent faults induced due to imperfections in the manufacturing process. 
Prior works, such as~\cite{zhang2018analyzing}, have shown that even a small fraction of these faults can drastically reduce the accuracy of DNNs. 
Hence, these faults render some of the fabricated chips useless, which negatively impacts the manufacturing yield. 
To address permanent faults in DNN hardware accelerators, various fault-mitigation techniques have been proposed. 
For example, Kim et al.~\cite{kim1989design} propose to bypass faulty Processing Elements (PEs) and view a faulty array as a smaller fault-free array. 
Such techniques improves the yield, but at the cost of performance loss. 
Techniques like~\cite{takanami2017built} propose to add redundancy such that each redundant PE in the architecture is dedicated for a limited region of the computing array. 
These techniques also significantly impact the performance of the system, as they employ redundancy for fault mitigation. 
Apart from redundancy-based techniques, \textit{Fault-Aware Pruning (FAP)}~\cite{zhang2018analyzing} is proposed which exploits intrinsic resilience of DNNs to pruning (zeroed weights/computations) to mitigate the effects of permanent faults in the computational array of a systolic array-based DNN accelerator. \textit{Fault-Aware Mapping (FAM)}~\cite{abdullah2020salvagednn} further improves the effectiveness of \textit{FAP} by permuting the DNN weights such that less significant weights are mapped to the bypassed (faulty/zeroed) PEs. 
The main shortcoming of these methods is that they offer fault-mitigation at the cost of accuracy loss. 
To offer fault mitigation without significant accuracy loss, \textit{Fault-Aware Pruning + Training (FAP+T)} is proposed in~\cite{zhang2018analyzing}. 
\textit{Fault-Aware Training (FAT)} is also exploited in~\cite{zhang2019fault} to mitigate permanent faults in DNN accelerators. 
The above works clearly show that \textit{FAT} leads to the best accuracy results under hardware faults, and it achieves this with minimal impact on system's performance. 
Even though \textit{FAT} offers the best accuracy, it has serious limitations. 
\textit{Its core drawback is that it incurs huge (re)training overheads, specifically in cases where a DNN has to be tuned for numerous faulty chips having distinct fault patterns. 
Towards this, we aim at addressing the following challenging question: how to reduce the (re)training overheads of FAT when a given DNN has to be tuned for numerous chips having different fault maps.}  

\subsection{Our Novel Contributions}

To address the above-mentioned research question, in this work, we present a novel framework, \textit{Reduce}. The framework mainly estimates the resilience of the given DNN to faults and defines the amount of retraining required for each individual faulty chip based on its fault characteristics and the resilience characteristics of the given DNN. 

\section{\textit{REDUCE}: Proposed Framework for Reducing the Overheads of Fault-Aware Retraining}
\label{Sec:proposed_framework}

\begin{figure}[t]
\centering
\includegraphics[width=1\linewidth]{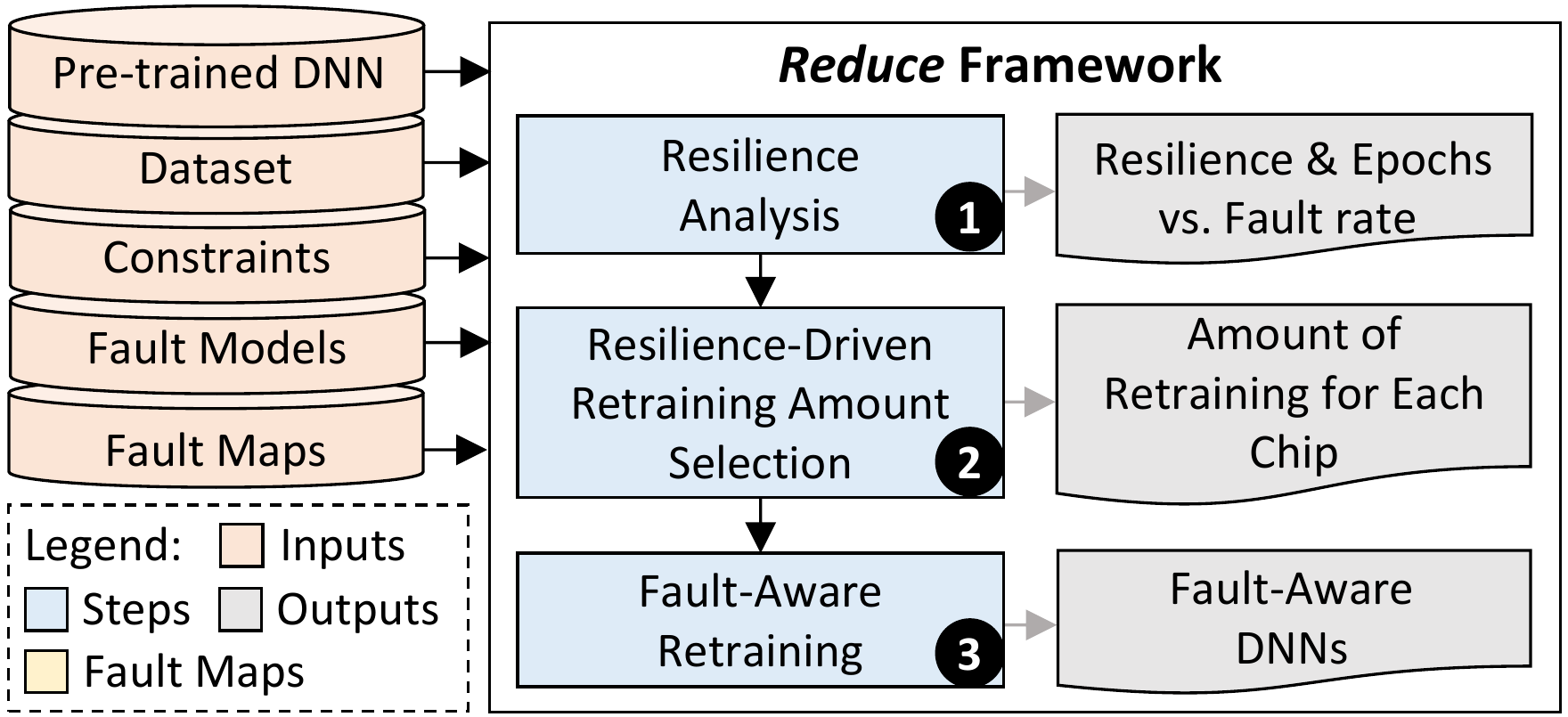}
\caption{Overview of the proposed \textit{Reduce} framework}
\label{fig:overall_framework}
\end{figure}

Fig.~\ref{fig:overall_framework} shows our proposed \textit{Reduce} framework, which receives a pre-trained DNN, a dataset, a user-defined accuracy constraint, and fault maps of the faulty chips as input and defines a retraining policy to efficiently generate fault-aware DNNs for the given faulty chips. 
The framework first computes the resilience of the given DNN to faults using fault-injection experiments at different fault rates and with different levels of retraining (Step~\circled{1}). 
This resilience is then used in Step~\circled{2} to select the amount of fault-aware retraining for each individual faulty chip based on its unique fault characteristics. 
The selection is performed in such a way that each output DNN offers accuracy close to the user-defined accuracy constraint without incurring unnecessary overheads. 
In the final step, Step~\circled{3}, \textit{FAT} is performed and the generated DNNs are then distributed to their corresponding faulty chips. 

\section{Results and Discussion}
\label{Sec:results}
\subsection{Experimental Setup}

To evaluate the effectiveness of the proposed technique, we consider the case of mitigating permanent faults in the computational array of a DNN accelerator.  
We consider the modified DNN accelerator design presented in~\cite{zhang2018analyzing} with FAP support. We assume the size of the systolic array to be $256 \times 256$. 
For evaluation, we built our entire framework using PyTorch. 
Similar to~\cite{zhang2018analyzing}, we consider a random fault injection model for generating fault maps. 

\subsection{Resilience Trends for VGG11 trained on Cifar10 Dataset}

\begin{figure}[hb]
\centering
\includegraphics[width=1\linewidth]{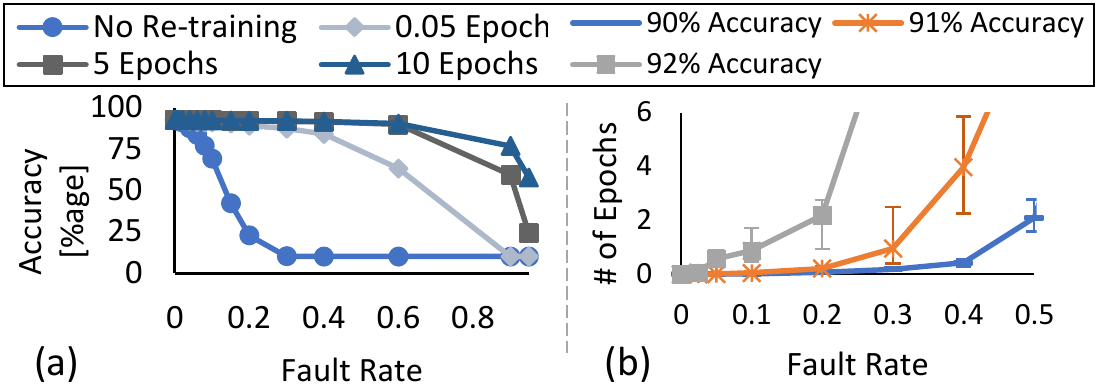}
\caption{Resilience trend of VGG11 trained on Cifar10 dataset.}
\label{fig:Resilience_plots}
\end{figure}

Fig.~\ref{fig:Resilience_plots}a shows the impact of different levels of FAT on the accuracy of the VGG11 at different fault rates while Fig.~\ref{fig:Resilience_plots}b shows the amount of FAT required at each fault rate to achieve a particular accuracy level. 
For each data point in Fig.~\ref{fig:Resilience_plots}b, we repeated the experiment five times and reported the minimum and maximum number of epochs along with the mean. 
The error bars in the figure show that the use of mean values can lead to undertraining. 
Therefore, we propose to use the maximum reported values for estimating the amount of retraining for each faulty chip, as it leads to higher confidence that the generated model meets the accuracy constraint. 

\subsection{Comparison with State of the Art}

\begin{figure}[t]
\centering
\includegraphics[width=1\linewidth]{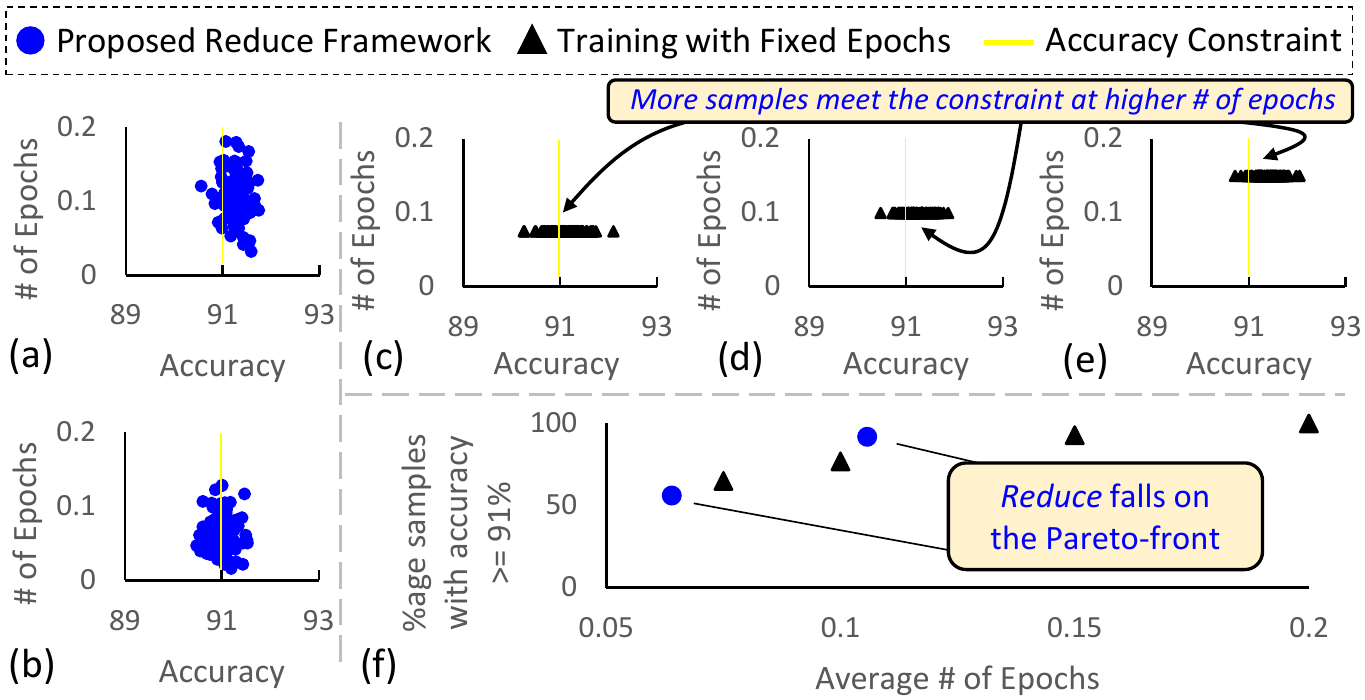}
\caption{Comparison with state of the art. (a) Results of the \textit{Reduce} framework using the maximum values from the resilience analysis for training amount estimation. (b) Results of \textit{Reduce} using mean values from the resilience analysis.
(c), (d) and (e) correspond to cases where VGG11 is trained for each chip using fixed number of epochs. (f) A summary of the results in (a)-(e). All the results are generated assuming 91\% as the accuracy constraint. }
\label{fig:SoA_comparison}
\end{figure}

To highlight the effectiveness of the proposed technique, we compared our \textit{Reduce} framework with fixed-policy retraining method proposed in~\cite{zhang2018analyzing}.  
Fig.~\ref{fig:SoA_comparison}a and~b shows the results of the proposed methodology when employed for retraining VGG11 (trained on Cifar10) for 100 faulty chips. 
Fig.~\ref{fig:SoA_comparison}c, \ref{fig:SoA_comparison}d and \ref{fig:SoA_comparison}e correspond to the cases where the DNN is trained for each faulty chip individually for a pre-specified number of epochs. 
The figures show that as the amount of retraining is increased the number of samples that meet the accuracy constraint increases. 
The results of Fig.~\ref{fig:SoA_comparison}a -~\ref{fig:SoA_comparison}e are summarized in Fig.~\ref{fig:SoA_comparison}f. 
The figures clearly show that the proposed \textit{Reduce} framework produces better (more robust) models with lesser training compared to the fixed-policy techniques. 

\section{Conclusion}
\label{Sec:conclusion}

In this paper, we proposed \textit{Reduce}, a methodology for reducing the overheads of fault-aware retraining when used for tuning a given DNN for multiple faulty chips. We mainly addressed how to compute the amount of retraining required for tuning the given DNN for a specific faulty chip such that the given DNN meets the user-defined accuracy constraint without incurring high retraining overheads. 
The results showed that the proposed technique could significantly reduce the retraining cost compared to state-of-the-art methods. 

\section*{Acknowledgement}
This work has been supported in part by the Center for Artificial Intelligence and Robotics (CAIR), funded by Tamkeen under the NYUAD Research Institute Award CG010, and the Center for Cyber Security (CCS), funded by Tamkeen under the NYUAD Research Institute Award G1104.

\def\bibfont{\footnotesize}
\bibliographystyle{IEEEtran}
\bibliography{biblio}

\end{document}